\documentstyle[12pt,epsf]{article}

\newcommand{\beq}{\begin{equation}}
\newcommand{\eeq}{\end{equation}}
\newcommand{\beqa}{\begin{eqnarray}}
\newcommand{\eeqa}{\end{eqnarray}}
\newcommand{\I}{{\rm i}}
\newcommand{\D}{{\rm d}}
\newcommand{\E}{{\rm e}}
\newcommand{\Z}{{\bf Z}}
\newcommand{\R}{{\bf R}}
\newcommand{\rr}{{\bf R}}
\newcommand{\Disp}{{\cal D}}
\newcommand{\Weyl}{{\cal W}}
\newcommand{\EC}{{\rm EC}}
\newcommand{\Nc}{N_{\rm c}}
\newcommand{\sgn}{\mathop{\rm sgn}\nolimits}
\newcommand{\Tr}{\mathop{\rm Tr}\nolimits}

\newcommand{\SUNc}{{\rm SU}(\Nc)}
\newcommand{\suNc}{{\rm su}(\Nc)}
\newcommand{\Dcov}{{\rm D}}
\newcommand{\der}[2]{\frac{\partial#1}{\partial#2}}

\newcommand{\ket}[1]{\mathopen|#1\rangle}

\newcommand{\ignore}[1]{}

\begin{document}

\begin{titlepage}
\begin{flushright}
FAU-TP3-96/15
\end{flushright}

\vfill

\begin{center}
{\LARGE The Path Integral for 1+1-dimensional QCD}

\vspace{1cm}

{\large O.~Jahn, T.~Kraus and M.~Seeger%
\footnote{\verb+mseeger@theorie3.physik.uni-erlangen.de+}}

{\em Institut f\"ur Theoretische Physik,\\
Friedrich-Alexander-Universit\"at Erlangen-N\"urnberg,\\
Staudtstra{\ss}e 7, 91058 Erlangen}
\end{center}

\vspace{2cm}

\begin{abstract}
We derive a path integral expression for the transition amplitude
in 1+1-di\-mensional QCD starting from canonically quantized QCD\@.
Gauge fixing after quantization \cite{LNT} leads to a formulation in terms
of gauge invariant but curvilinear variables. Remainders of
the curved space are Jacobians, an effective potential,  
and sign factors just as for the problem of a particle in a box.
Based on this result we derive a Faddeev-Popov like expression for the
transition amplitude avoiding standard infinities that are caused
by integrations over gauge equivalent configurations.
\end{abstract}

\vspace{5cm}
\end{titlepage}

\section{Introduction}
The standard Faddeev-Popov path integral formulation \cite{Itzykson}
provides a convenient framework for calculations in the
perturbative regime of QCD\@.  There are, however, some points of
interest that need clarification:

\begin{itemize}
\item
The derivation of the Faddeev-Popov path integral involves infinities
from the integration over gauge equivalent configurations.  Therefore
it would be preferable to find a derivation of a QCD path integral
which is well-defined at all stages.
\item The usual Faddeev-Popov formulation in the continuum gives no
prescription for how to handle infrared singularities, which may play
an essential role for the non-perturbative aspects of QCD\@.
\item There is a lively discussion about the detailed form
of the Faddeev-Popov path integral, concerning issues like
Jacobian factors and the appearance of the Haar measure 
\cite{LNT,LST,Reinhardt,Hetrick,Hosotani}.
\end{itemize}

In order to clarify these issues we choose canonically quantized QCD
as a well-defined starting point.
By working in 1+1 dimensions we avoid field theoretical infinities
indicating the need for regularization; only infinities resulting from
gauge invariance are left. 
We derive a path integral expression for the transition amplitude
in the traditional manner by inserting eigenstates of the gauge field
operators and coherent states for the fermions respectively. 
With the help of a gauge fixing procedure following Lenz et~al.~\cite{LNT},
we achieve  formulations purely in terms of physical (i.e.\ unconstrained)
but curvilinear variables. We derive two equivalent expressions 
for the path integral depending on the domain of definition (compact
vs.\ non-compact) of the remnants of the gauge fields.
The inclusion of fermions entails no further difficulties and is treated
in the appendix.
As an illustration of the formalism we explicitly calculate the
partition function and the spectrum of pure SU($\Nc$) Yang-Mills theory.
Finally we establish the connection to the  Faddeev-Popov
formalism avoiding the usual infinities.

To make the paper self-contained and to fix our notation we first
summarize the formalism.

\section{Review of the Canonical Formalism and Conventions}

In this section we give a short review of the formulation of 1+1-dimensional
Yang-Mills theory in the canonical formalism, based on the work of
Lenz, Naus and Thies~\cite{LNT}.

The Hamiltonian of 1+1-dimensional Yang-Mills theory in the Weyl
gauge is given by
\begin{equation}
H = \int_0^L \D x \Tr \bigl( \Pi(x) \bigr)^2 ,
\label{Weyl-Hamiltonian}
\end{equation}
where $\Pi(x)$ is the momentum canonically conjugate to $A(x)$, and
equals, up to a sign, the chromo-electric field.  $A(x)$ and $\Pi(x)$
are $\suNc$ matrices, that can be expanded in terms of the $\Nc^2-1$
standard generators $\lambda^a/2$,
\begin{equation}
A(x) = A^a(x) \frac{\lambda^a}{2} \quad \mbox{and} \quad
\Pi(x) = \Pi^a(x) \frac{\lambda^a}{2} .
\end{equation}
(Repeated indices are summed over throughout the paper.)  Space is
assumed to be a circle, i.e., periodic boundary conditions
$A(x+L)=A(x)$ and $\Pi(x+L)=\Pi(x)$ are imposed.  The theory is
quantized by demanding the commutation relations
\begin{equation}
[\Pi^a(x),A^b(y)] = -\I \delta^{ab} \delta(x-y) ,
\end{equation}
where $\delta(x-y)$ denotes a periodic $\delta$-function.

In the Weyl gauge, Gau\ss's law has to be required as a constraint on
physical states,
\begin{equation}
\Dcov\Pi \ket{\Phi} = 0 , \qquad \mbox{where }
\Dcov^{ab} = \delta^{ab}\der{}{x} + g f^{acb} A^c .
\end{equation}
Since the operator $\Dcov\Pi$ generates small gauge transformations,
this means that physical wave functions have to be gauge invariant.
(In 1+1 dimensions there are no large gauge transformations because
$\pi_1(\SUNc)$ is trivial\footnote{ Strictly speaking, in the absence
of fermions in the fundamental representation the gauge group is
$\SUNc/\Z_{\Nc}$, which has a non-trivial fundamental group, but since
we are going to include fundamental fermions later, we ignore this
fact.}.)

The Gau\ss{} law constraint is solved in~\cite{LNT} by a transition to
curvilinear coordinates $(a,A')$, where the coordinates $a$ constitute
a maximal set of gauge invariant quantities, while $A'$ represents the
gauge variant part of $A$.  Thus, physical wave functions are
functions of $a$ only.  More specifically, $a$ is defined as a
constant diagonal matrix containing the phases of the eigenvalues of
the Wilson loop around the circle:
\begin{equation}
\E^{\I g L a} \equiv V^{\dagger} {\cal P}
\exp \left( \I g \int_0^L \D x \, A(x) \right) V ,
\label{diag}
\end{equation}
where $\cal P$ denotes path ordering and $V\in\SUNc$ is chosen such that
\begin{equation}
a = {\rm diag} (a_1, \dots, a_{N_{\rm c}}) 
= a^{c_0}\frac{\lambda^{c_0}}{2} .
\label{adiag}
\end{equation}
Because only $\Nc-1$ of the $\Nc$ variables $a_i$ are independent, $a$
is expanded in terms of the $\Nc-1$ generators of the Cartan
subalgebra of $\SUNc$ in the last equation\footnote{The index $c_0$
takes on the values $i^2-1$ with $i=2,\ldots,\Nc$ throughout the
paper.}.

\begin{figure}[ht]
\centerline{\epsffile{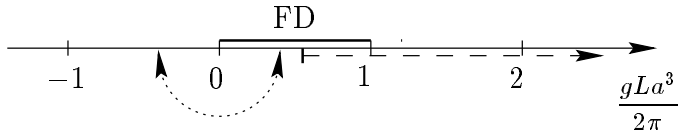}}
\caption{Fundamental domain for SU(2).  FD denotes the fundamental
	domain, the dashed arrow is a displacement, the dotted arrow
	(a reflection at the origin) is a color permutation}
\label{fund2} 
\end{figure}

\begin{figure}[ht]
\centerline{\epsffile{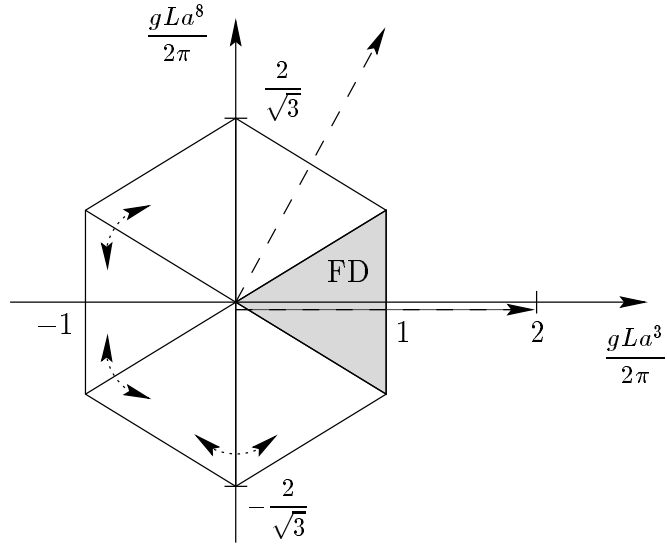}}
\caption[Fundamental domain for SU(3)]{Fundamental domain for SU(3).
	The shaded area is the fundamental domain, the dashed arrows
	are displacements and the dotted arrows (reflections at the
	diagonals of the hexagon) are color permutations}
\label {fund3}
\end{figure}

As can be seen from (\ref{diag}), the variables $a$ are defined only
up to
\begin{eqnarray}
&\bullet& \mbox{displacements} \quad
a_i \longrightarrow a_i + \frac{2\pi}{gL} k_i \quad
\mbox{for } k_i \in \Z \mbox{ with } \sum_i k_i = 0 \quad 
\label{displacements} \\
&\bullet& \mbox{and color permutations} \quad
a_i \longrightarrow a_{P(i)} \quad
\mbox{for } P \in {\rm S}_{\Nc} , \quad 
\label{permutations}
\end{eqnarray}
where S$_{\Nc}$ denotes the symmetric group.  The set of all
displacements will sometimes be denoted by $\Disp$, the set of all
color permutations by $\Weyl$ (the `Weyl group' of $\SUNc$).  Examples
for the cases of SU(2) and SU(3) are shown in figs.~\ref{fund2}
and~\ref{fund3}, resp.

The invariant hyperplanes of all reflections contained in
$\Disp\times\Weyl$ divide $\R^{\Nc-1}$ into domains in such a way
that any $a\in\R^{\Nc-1}$ can be mapped into a given domain by
exactly one transformation in $\Disp\times\Weyl$.  To render the
definition of $a$ unique, one can therefore restrict its values to one
of the domains, which is then called the {\em fundamental domain}
(FD)\@.  A special choice for SU(2) and SU(3) is shown in
figs.~\ref{fund2} and~\ref{fund3}.

The remaining coordinates $A'$ can be chosen such that the Jacobian of
the coordinate transformation factorizes,
\begin{equation}
\der{(a,A')}{(A)} = J(a) J'[A'] ,
\end{equation}
where $J(a)$ results only from the diagonalization in~(\ref{diag}) and
equals the density of the `radial' (class) part of the Haar measure of
$\SUNc$,
\begin{equation}
J(a) = \prod_{i<j} \sin^2 \left( \frac{gL(a_i-a_j)}{2} \right) .
\label{jacobian}
\end{equation}

In the new coordinates the Hamiltonian~(\ref{Weyl-Hamiltonian}),
when acting on physical states, reduces to the radial part of the
Laplace-Beltrami operator of $\SUNc$,
\begin{eqnarray}
H \ket{\Phi} &=& H_{\rm ph} \ket{\Phi} , \nonumber\\
H_{\rm ph} &\equiv& - \frac{1}{2L} \frac{1}{J(a)} \der{}{a^{c_0}} J(a) 
\der{}{a^{c_0}} .
\label{yangmillsphysical}
\end{eqnarray}

We are now in a position to formulate the path integral for this problem.
We will derive two forms with different domains of $a$:
\begin{enumerate}
\item $a$ is defined on the covering space ${\R}^{N_{\rm c}-1}$.
\item $a$ is restricted to the FD and all integrations are performed
over the FD.
\end{enumerate}
In the first case the transformations (\ref{displacements}) and
(\ref{permutations}) will be considered as residual symmetries,
under which physical wave functions have to be invariant.

\section{Derivation of the Path Integral on the Covering Space}
\label{covering space}

In this section we derive a path integral expression for the time
evolution of a physical state $|\Phi\rangle$, i.e.\ for the amplitude
$\langle A | \Phi (t) \rangle$, on the covering space.

As usual, we divide the time interval into a large number $N$ of time
slices, each of length $\varepsilon = t/N$ and insert unity at each
step,
\beq
\langle A | \,\E^{-\I H t} | \Phi \rangle = \int \prod_{i=1}^{N}
\left[ \D A_{i-1} \, \langle A_{i}| \,\E^{-\I H_{\rm
ph}\varepsilon}|A_{i-1}
\rangle \right] \langle A_0 | \Phi \rangle, \label{amplitude}
\eeq
with the identification $A_N \equiv A$.
In (\ref{amplitude}) $H$ has been replaced by $H_{\rm ph}$ using
(\ref{yangmillsphysical}) and the fact that $H$ commutes with the
projection onto physical states.

\begin{sloppypar}
The matrix elements for the infinitesimal time evolution are now
evaluated in a Schr\"odinger representation where we pay attention to
the fact that the coordinate transformation yielded a non-trivial
Jacobian,
\beqa
\lefteqn{\langle A_{i}| \,\E^{-\I H_{\rm ph} \varepsilon} | A_{i-1} \rangle = \exp
\left[ \I \varepsilon \frac {1}{2 L} \left( \frac
{\partial^2}{\partial {a_i^{c_0}}^2} + f_{c_0}(a_i)
\frac{\partial}{\partial a_i^{c_0}}\right) \right]\times{}} \nonumber \\
& & \quad\quad\quad\quad\quad\quad\quad\quad
{}\times \frac{\delta (a_{i}-a_{i-1})}{\sqrt{J(a_{i})J(a_{i-1})}}
\frac{\delta [A'_{i}-A'_{i-1}]}{\sqrt{J'[A'_{i}]J'[A'_{i-1}]}}.
\eeqa
The Jacobians have been split in such a way that we can obtain a
simpler form of the kinetic energy by moving one of the factors
through the differential operators.  But before this we have to ensure
differentiability of the terms the operators act on, which can be
achieved by use of the identity
\beq
\frac{\delta (a_{i}-a_{i-1})}{\sqrt{J(a_{i})J(a_{i-1})}} =
\frac{\sgn a_{i} }{\sqrt{J(a_{i})}}
\, \frac{\sgn a_{i-1}}{\sqrt{J(a_{i-1})}}
\, \delta (a_i-a_{i-1}),
\eeq
where $\sgn a$ equals $+1$ if an even number of reflections is needed
to map $a$ into the FD and $-1$ otherwise.  This sign makes the root
of the Jacobian differentiable:
\beq
\sgn a \, \sqrt{J(a)} =
\sgn a \, \prod_{k<l} \left| \sin \frac{gL(a_l-a_k)}{2}  
\right|
\propto \prod_{k<l} \left( \sin \frac{gL(a_l-a_k)}{2}  \right).
\eeq
\end{sloppypar}

If we now move the Jacobian factors to the left, the kinetic
energy simplifies and an effective potential is induced,
\beq
\exp \left[ \I \varepsilon \frac {1}{2 L} \left( \frac
{\partial^2}{\partial {a^{c_0}}^2} + f_{c_0}(a)
\frac{\partial}{\partial a^{c_0}} \right) \right]
\frac{\sgn a }{\sqrt{J(a)}} 
= \frac{\sgn a}{\sqrt{J(a)}} \exp \left[
\I \varepsilon \left( \frac {1}{2 L}
\frac {\partial^2}{\partial {a^{c_0}}^2} - V_{\rm eff}
\right) \right],
\eeq
which is of course reminiscent of the hydrogen atom where $V_{\rm
eff}$ would be the centrifugal barrier. The procedure just applied
corresponds to the transition to ``reduced'' wave functions.
Here, however, $V_{\rm eff}$
can be calculated to yield an irrelevant constant \cite{LNT}. It follows
that
\beqa
\lefteqn{\langle A_i| \,\E^{-\I H_{\rm ph} \varepsilon} | A_{i-1} \rangle
=\frac{\delta
[A'_i-A'_{i-1}]\cdot \sgn a_i \cdot \sgn a_{i-1}}
{\sqrt{J(a_i)J(a_{i-1})J'[A'_i]J'[A'_{i-1}]}}
\,(2\pi)^{1-N_{\rm c}} \times{}} \nonumber \\
& &\quad\quad\quad\quad\quad\quad {}\times \int\limits_{\rr^{N_{\rm c}-1}}
 \D p_i \, \exp \left[ \I \varepsilon
\left( p_i \frac{a_i-a_{i-1}}{\varepsilon} - \frac{1}{2L} p_i^2 - V_{\rm
eff} \right) \right],
\label{amatrixelement}
\eeqa
where the integral representation of the $\delta$-function has been
inserted and the index $c_0$ has been supressed from $p$ and $a$.

Let us now consider the projection
onto physical states. Simply requiring them to be
independent of the $A'$ with $a$ varying over the whole $\R^{N_{\rm
c}-1}$ is not sufficient to render them physical. In addition, we have
to postulate invariance under displacements and color permutations.
A physical state would therefore be, e.g.
\beq
| \Phi\rangle = | a \rangle_{\rm phys} 
\equiv \sum_{\omega\in \Weyl\times \Disp}
	| {}^{\omega}a \rangle.
\eeq
After insertion into (\ref{amplitude}) this leads to the phase
space path integral\footnote{In our final expression for the path integral
we perform the projection onto physical states at the endpoint. It is
irrelevant at which time the projection is done, because the Hamilton
operator is itself invariant under displacements and permutations.}
\beqa
\lefteqn{\langle a_{\rm f}| \,\E^{-\I H_{\rm ph} t} | a_{\rm i}
\rangle_{\rm phys} = \frac{1}{\sqrt{J(a_{\rm f})J(a_{\rm i})}} \sum_{\omega}
\sgn \omega \int\limits_{\rr^{N_{\rm c}-1}}
 \prod_{i=1}^{N-1} \D a_i \, (2\pi)^{(1-N_{\rm c})N}
\times{}} \nonumber \\
& & {}\times \int\limits_{\rr^{N_{\rm c}-1}}
 \prod_{i=1}^{N} \D p_i \, \exp \left[ \I
\varepsilon \sum_{i=1}^N \left( p_i 
\frac{a_i-a_{i-1}}{\varepsilon} - \frac{1}{2L} p_i^2 - V_{\rm eff}
\right) \right] \Biggr|_{{\textstyle a}_0={\textstyle a}_{\rm
i}}^{{\textstyle a}_N={\textstyle {}^{\omega}a}_{\rm f}},
\label{phase}
\eeqa
where we have assumed that $a_{\rm f}, a_{\rm i} \in {\rm FD}$ and
have performed the integration over the unphysical degrees of
freedom with help of
\beq
\int \D A \, (.) = \int \D a \, J(a) \int \D A' \, J'[A'] \, (.).
\eeq
In eq.~(\ref {phase}) only the $\sgn \omega$ factors of the endpoints
are left, while all others have canceled pairwise.
The configuration space path integral is obtained by performing the
Gaussian integrations over the momenta,
\beqa
\lefteqn{\langle a_{\rm f}| \,\E^{-\I H_{\rm ph} t} | a_{\rm i}
\rangle_{\rm phys} = \frac{1}{\sqrt{J(a_{\rm f})J(a_{\rm i})}} \left(
\frac{L}{2\pi \I \varepsilon} \right)^{(N_{\rm c}-1)N/2} \sum_{\omega} \sgn \omega
\times{}} \nonumber \\
& & {}\times \int\limits_{\rr^{N_{\rm c}-1}} \prod_{i=1}^{N-1} \D a_i \, \exp \left[ \I
\varepsilon \sum_{i=1}^N \left(
\frac{L}{2}\frac{(a_i-a_{i-1})^2}{\varepsilon^2} - V_{\rm eff} \right)
\right] \Biggr|_{{\textstyle a}_0={\textstyle a}_{\rm i}}^{{{\textstyle
a}_N={\textstyle{}^{\omega}a}_{\rm f}}},
\label{config1}
\eeqa
with  $a_{\rm f}, a_{\rm i} \in {\rm FD}$.

The inclusion of Fermions can be accomplished along similar lines and
is therefore postponed to the appendix.

\section{Derivation of the Path Integral on the Fundamental Domain}
We start with the Schr\"odinger equation for the reduced wave
function $\Psi(a) = \sqrt{J(a)} \, \Phi(a)$:
\beq
\left(- \frac {1}{2 L}
\frac {\partial^2}{\partial {a^{c_0}}^2} + V_{\rm eff} 
\right) \Psi(a) = {E} \Psi(a),
\label{schroe}
\eeq
together with the condition $\Psi(a)=0$ if $J(a) = 0$. 
 Because the reduced wave function has to vanish on the
boundary of the FD, the solution of the Schr\"odinger equation 
(\ref {schroe}) in  the FD is not influenced by  
the solutions in its gauge copies.

We now reformulate the problem. We enforce the vanishing of the
wave function on the boundary of the FD by introducing a potential,
which is equal to zero inside and infinite outside the FD\@.
Thus we obtain the new Schr\"odinger equation
\beq
\left(- \frac {1}{2 L}
\frac {\partial^2}{\partial {a^{c_0}}^2} + V_{\rm eff} 
+ V_{\rm box}(a)
\right) \Psi(a) = {E} \Psi(a),
\label {newschroe}
\eeq
with  \[  V_{\rm box} (a)  = \left\{ \begin{array}{r@{\quad:\quad}l}
       0 & a \in {\rm FD} \\
       \infty & {\rm otherwise,}
       \end {array} \right. \]
and no  constraint on the reduced wave function $\Psi$.

As we have a standard kinetic energy and a potential which is bounded 
from below we can apply the Trotter product formula \cite{Roepstorff}
\beq
\E^{-\I H_{\rm ph} t} = \lim_{N \rightarrow \infty} \left(\E^{-\I
\left(V_{\rm eff} + V_{\rm box}(a) \right) t/N } \cdot \E^{-\I E_{\rm
kin} t/N}\right)^{N},
\eeq
and then derive the path integral in the usual way.
To be consistent with the previous section we consider the reduced
$a$-eigenstates $\sqrt{J(a)} \, | a \rangle$ with the property
\beq
\sqrt{J(a')} \, \langle a' | a \rangle \sqrt{J(a)} = \delta (a - a') .
\eeq
We obtain
\beqa
\lefteqn{
\langle a_{\rm f}| \,\E^{-\I H_{\rm box} t} | a_{\rm i} \rangle 
 =  \frac{1}{\sqrt{J(a_{\rm f})J(a_{\rm i})}} \,
 \int\limits_{\rr^{N_{\rm c}-1}} \prod_{i=1}^{N-1} \D a_i \,
(2\pi)^{(1-N_{\rm c})N} \, \int\limits_{\rr^{N_{\rm c}-1}}
\prod_{i=1}^{N} \D p_i \times{}} \nonumber \\
& &\quad \quad \quad \quad  \quad   \left. {}\times  \, \exp \left[ \I
\varepsilon \sum_{i=1}^N \left( p_i \frac{a_{i}-a_{i-1}}{\varepsilon}
- \frac{1}{2L} p_i^2 - V_{\rm eff} - V_{\rm box} (a_i) \right) \right]
\right|_{{\textstyle a}_0 = {\textstyle a}_{\rm i}}^{{\textstyle
a}_N={\textstyle a}_{\rm f}}. \nonumber 
\label{phase2}
\eeqa
Due to the potential $V_{\rm box}$, which is rapidly oscillating
outside the FD, the $a$-integration can be restricted to give the
phase space path integral on the FD
\beqa
\lefteqn{
\langle a_{\rm f}| \,\E^{-\I H_{\rm box} t} | a_{\rm i} \rangle 
 =  \frac{1}{\sqrt{J(a_{\rm f})J(a_{\rm i})}} \,
 \int\limits_{\rm FD} \prod_{i=1}^{N-1} \D a_i \, (2\pi)^{(1-N_{\rm c})N}
\, \int\limits_{\rr^{N_{\rm c}-1}} \prod_{i=1}^{N} \D p_i \times{}} \nonumber \\
& &\quad \quad \quad \quad  \quad \quad\quad\quad\quad \left. {}\times  \,
\exp \left[ \I \varepsilon \sum_{i=1}^N \left( p_i
\frac{a_{i}-a_{i-1}}{\varepsilon} - \frac{1}{2L} p_i^2 - V_{\rm eff}
\right) \right] \right|_{{\textstyle a}_0={\textstyle a}_{\rm
i}}^{{\textstyle a}_N={\textstyle a}_{\rm f}}.
\nonumber \\ &&
\eeqa
The configuration space path integral is obtained by performing the
Gaussian integrations over the momenta,
\beqa
\lefteqn{\langle a_{\rm f}| \,\E^{-\I H_{\rm box} t} |
a_{\rm i}
\rangle   =   \frac{1}{\sqrt{J(a_{\rm f})J(a_{\rm i})}} \left(
\frac{L}{2\pi \I \varepsilon} \right)^{(N_{\rm c}-1)N/2} 
\times{}} \nonumber \\
& & \quad\quad\quad\quad\quad\quad\quad
{}\times \int\limits_{\rm FD} \prod_{i=1}^{N-1} \D a_i \, \exp \left[ \I
\varepsilon \sum_{i=1}^N \left(
\frac{L}{2}\frac{(a_{i}-a_{i-1})^2}{\varepsilon^2} - V_{\rm eff}
\right)
\right] \Biggr|_{{\textstyle a}_0={\textstyle a}_{\rm
i}}^{{\textstyle a}_N={\textstyle a}_{\rm f}}.
\nonumber \\ &&
\label{config2}
\eeqa

\section{Equivalence of the Two Formulations}
In the previous sections with have derived two different formulations
of the path integral for 1+1-dimensional Yang-Mills theory.
We will show now that they
are equivalent in the  limit $\varepsilon \rightarrow 0$, 
where we deal with paths which are continuous in time.
In the {\it formulation on the covering space} we then have a sum over
endpoints, which are related by small gauge transformations and are
accompanied by a possible minus sign.
We demonstrate that only paths which stay within the FD are relevant
because all paths that touch the boundary or cross it cancel each
other.

To calculate the transition amplitude
$\langle a_{\rm f}| \,\E^{-\I H_{\rm ph} t} | a_{\rm i} \rangle_{\rm phys}$
we have to sum over all paths 
starting at $a(0)=a_{\rm i}$ and ending at $a(t)={}^\omega a_{\rm f}$
including a sum over the symmetry transformations $\omega$.

\begin{figure}[hhh]
\centerline{\epsffile{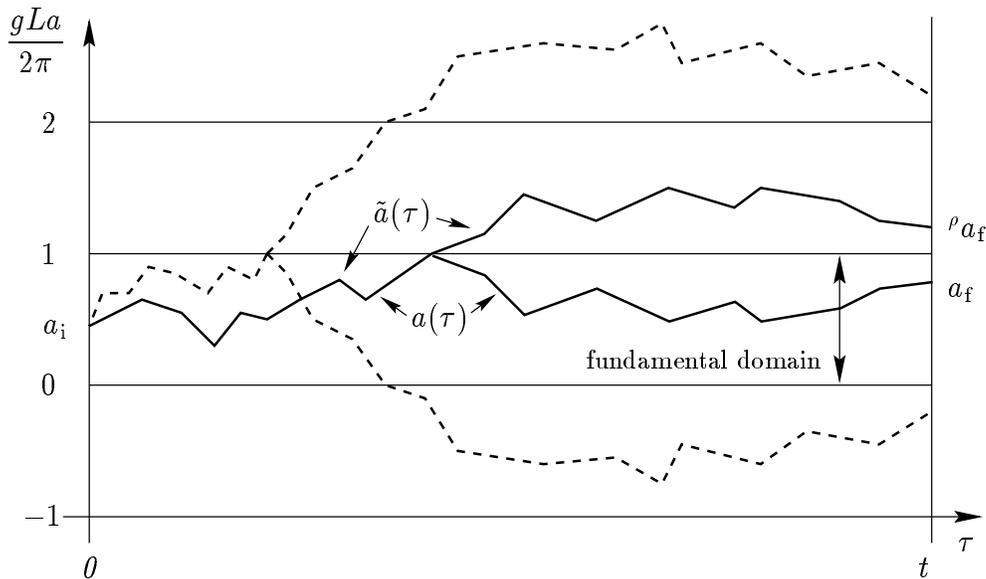}}
\caption [Cancellation of Paths]
           {\label {path} Cancellation of Paths  }
\end{figure}

Let $t_1 = \inf \{ \tau\,|\, 0 < \tau <t,\, a(\tau) \in \partial{\rm
FD}\}$ be the first time that $a(\tau)$ touches the boundary of the FD\@.
Now consider the path
 \[  \tilde a(\tau)  = \left\{ \begin{array}{r@{\quad:\quad}l}
       a(\tau) & 0 \leq \tau \leq t_1 \\
       {}^\rho a(\tau) & t_1 \leq \tau \leq t,
       \end {array} \right. \]
where $\rho \in \Weyl \times \Disp$ denotes the reflection at the
hyperplane  the path $a(\tau)$ meets at $\tau=t_1$. 
 Because the reflection at the boundary is a symmetry of the
Yang-Mills Hamiltonian, the two paths $\tilde a(\tau)$  and
$a(\tau)$ have the same action\footnote{This argument holds also for
the inclusion of fermions, if all gauge transformations of the $a$
are accompanied by the appropriate gauge transformations of the
fermions.}.
But as the final points of the two paths differ by a reflection,
they have opposite sign and cancel each other in the sum over symmetry
transformations in eqs.~(\ref{phase}) and~(\ref {config1}).
Thus it is equivalent to consider only paths which stay inside the FD and
we have arrived at the second form (\ref{phase2}, \ref{config2}) of the
path integral.

For the case of ${\rm SU}(2)$ the cancellation of paths is illustrated 
in fig.~\ref{path}.

\section{Explicit Calculation of the Transition Amplitude and the 
	 Partition Function}\label{explicit}

To illuminate the role of the sum over endpoints
we derive an explicit expression for the transition amplitude
(\ref{config1}) in terms of energy eigenstates.  The Gaussian
integrations in (\ref{config1}) can be readily performed to yield the
well-known propagator of a free particle plus a sum over endpoints,
a sign and a Jacobian factor: 
\beq
\langle a_{\rm f}| \,\E^{-\I H_{\rm ph} t} | a_{\rm i}
\rangle_{\rm phys} 
= \frac{1}{\sqrt{J(a_{\rm f})J(a_{\rm i})}} 
\left( \frac{L}{2\pi \I t} \right)^{(\Nc-1)/2} 
\sum_{\omega} \sgn \omega
\exp \left[ \I \frac{L}{2t} ({}^{\omega}a_{\rm f}-a_{\rm i})^2 
- \I V_{\rm eff} t \right] .
\label{expl1}
\eeq 
Now we split the sum over $\omega$ into a sum over color permutations
$\rho\in\Weyl$ and a sum over displacements $\delta\in\Disp$.
The sum over displacements is turned into a sum over momenta by use of
Poisson's resummation formula,
\beq
\sum_{\delta\in\Disp} f(a+\delta) = \frac{1}{|\EC|} \sum_{p\in\Disp^*}
\, \int\limits_{\rr^{\Nc-1}} \!\! \D a' \, \E^{-\I p a'} f(a+a') ,
\label{poisson}
\eeq
where $|\EC|$
is the volume of the elementary cell (the Wigner-Seitz cell of $\Disp$)
and $\Disp^*$ denotes the reciprocal lattice of the Bravais lattice
$\Disp$. It is defined by
\beq
p \delta = 2 \sum_{i=1}^{\Nc} p_i \delta_i \in 2\pi\Z
\qquad\mbox{for } \delta_i\in\frac{2\pi}{gL}\Z,\ \sum_i \delta_i=0,
\eeq
or, since $\sum_i p_i=0$,
\beq
p_i = \frac{gL}{2} \left( n_i - \frac{1}{\Nc} \sum_i n_i \right)
\qquad\mbox{with } n_i\in\Z,\ n_{\Nc}=0.
\label{reclatt}
\eeq
Here, following the convention of (\ref{adiag}), $p_i$ and $\delta_i$
denote the diagonal matrix elements of $p$ and $\delta$, resp.

Applying (\ref{poisson}) to (\ref{expl1}) and performing the
$a'$-integration we can write
\beq
\langle a_{\rm f}| \,\E^{-\I H_{\rm ph} t} | a_{\rm i} \rangle_{\rm phys} 
= \frac{1}{\sqrt{J(a_{\rm f})J(a_{\rm i})}\,|\EC|} 
\sum_{p\in\Disp^*} \sum_{\rho\in\Weyl} \sgn \rho
\exp \left[ - \I \frac{t}{2L} p^2 + \I p ({}^{\rho}a_{\rm f}-a_{\rm i}) 
- \I V_{\rm eff} t \right] .
\eeq

The $p$ that are invariant under some permutation $\rho\in\Weyl$,
and therefore also under some transposition $\tau$, do not contribute
because for these the exponential is invariant under the substitution
$\rho\to\tau\rho$ and therefore each term with a given $\rho$ is
canceled by the term with $\tau\rho$.
For the other $p$ we choose out of each multiplet of $\Nc!$
vectors the one with $p_{1}>p_{2}>\ldots>p_{\Nc}$ and sum over the
permuted ones separately.  After rearranging the $\rho$-sum we obtain
\beqa
\lefteqn{\langle a_{\rm f}| \,\E^{-\I H_{\rm ph} t} | a_{\rm i}
\rangle_{\rm phys} 
= \frac{1}{\sqrt{J(a_{\rm f})J(a_{\rm i})}\,|\EC|} 
\sum_{\textstyle{p\in\Disp^* \atop p_{1}>\ldots>p_{\Nc}}}
\sum_{\rho,\sigma\in\Weyl} \sgn(\rho\sigma)
\times{}} \qquad\qquad\qquad\qquad \nonumber \\
& & {}\times \exp \left[ - \I \left( \frac{1}{2L} p^2
+ V_{\rm eff} \right) t \right]
\E^{- \I p{}^{\rho}\!a_{\rm f}} \E^{\I p{}^{\sigma}\!a_{\rm i}} .
\label{expl2}
\eeqa

This formula provides us immediately with the energy spectrum of
1+1-dimensional Yang-Mills theory.  Using the representation
(\ref{reclatt}) and the expression for $V_{\rm eff}$ found in~\cite{LNT}
one obtains
\beq
E_n = \frac{g^2L}{4} \left[ \sum_{i=1}^{\Nc-1}n_i^2 - \frac{1}{\Nc}
\left( \sum_{i=1}^{\Nc-1} n_i \right)^2 - \frac{\Nc(\Nc^2-1)}{12} \right]
\eeq
with $n_1>n_2>\ldots>n_{\Nc-1}>0$.  For SU(2) this reads
\beq
E_n = \frac{g^2L}{8} (n_1^2-1)
\eeq
and for SU(3)
\beq
E_n = \frac{g^2L}{6} (n_1^2 + n_2^2 - n_1 n_2 - 3) .
\eeq
The corresponding eigenfunctions can also be read off:
\beq
\Phi_p(a) = \frac{1}{\sqrt{J(a)|\EC|}} 
\sum_{\rho\in\Weyl} \sgn \rho \, \E^{\I p {}^{\rho}\!a} .
\eeq
These are just the characters of the irreducible representations of
SU($\Nc$).  We see that the sum over color permutations together with
the sign ensures that the zeros of the Jacobian are canceled and the
wave functions remain finite.

Our results agree with those of Rajeev~\cite{Rajeev}, Gupta
et~al.~\cite{Gupta}, Engelhardt~\cite{Engelhardt}, Engelhardt and
Schreiber~\cite{Schreiber} and the compact
formulation of Hetrick~\cite{Hetrick} but disagree with the
non-compact formulation appearing in the latter.  In this formulation
the gauge is fixed on the classical level and the resulting
configuration space, the Cartan subalgebra of SU($\Nc$) modulo the
Weyl group, is quantized under the assumption of a flat measure.  As
opposed to this, in Rajeev's as well as in our (resp.~\cite{LNT}'s)
approach the full constrained system is quantized, and in the course
of implementing the constraints the configuration space turns out to
be the set of classes of SU($\Nc$) (the eigenvalues of the Wilson
loop) equipped with the Haar measure, which is induced by the flat
measure of the original (unconstrained) configuration space.

In an earlier work \cite{Hosotani} Hetrick and Hosotani derive the
non-compact results also via a Faddeev-Popov path integral and find
that the Haar measure (the Faddeev-Popov determinant) is canceled by
the $A^0$-integration.  In section \ref{Faddeev-Popov} we will show
that our result is equivalent to a Faddeev-Popov-like expression with
an additional antisymmetrizing sum over gauge copies and that the Haar
measure is indeed canceled.
The calculation in this section shows that these eliminate that part
of the spectrum that differs between their and our approach.

The partition function
\beq
Z(\beta) = \Tr\E^{-\beta H} = \int_{\rm FD} \D a\, J(a)\, 
\langle a| \,\E^{-\beta H_{\rm ph}} | a\rangle_{\rm phys} ,
\label{part1}
\eeq
where the trace is taken only over the physical part of the Hilbert
space, contains a matrix element that differs from the transition
amplitude calculated in section \ref{covering space} only by a factor of
$\I$ in the exponent.  The derivation of (\ref{config1}) and
(\ref{expl2}) is still valid after the replacement of $\I t$ by
$\beta$ and $\I\varepsilon$ by $\varepsilon$, and so we obtain,
after inserting (\ref{expl2}) into (\ref{part1}) and performing the
integration,
\beq
Z(\beta) = \sum_{\textstyle{p\in\Disp^* \atop p_{1}>\ldots>p_{\Nc}}}
\exp \left[ - \beta \left( \frac{1}{2L} p^2 + V_{\rm eff} \right)\right].
\eeq

\section{Connection with the Faddeev-Popov Formalism}
\label{Faddeev-Popov}

Having derived the gauge fixed path integral, one is tempted to ask
for the connection between this approach and the more conventional
Faddeev-Popov approach. In this section we show that the two
expressions for the partition function are actually equivalent if the
infrared degrees of freedom in the Faddeev-Popov path integral are
treated correctly.  To this end we continue the transition
amplitude~(\ref{config1}) with endpoints identified to Euclidean space
and insert the result into~(\ref{part1}):
\beqa 
Z(\beta) = \int_{\rm FD} \D a^1(0)\, \sum_{\omega} \sgn\omega
\int_{a^1(0)}^{^{\omega}\!a^1(0)} {\cal D}a^1
\exp \left[ - \int_0^{\beta} \D \tau \left( \frac{L}{2} (\dot{a}^1)^2
- V_{\rm eff} \right) \right]
\label{cont.PI}
\eeqa
Here we have adopted the usual continuum notation for the path
integral measure and integrand.  In addition, we have reintroduced
the spatial Lorentz index on $a$.

The axial gauge corresponds to the gauge condition $f[A] = 0$ with
\beq
f^{c_0}[A] = \partial_1 A^{1c_0} + \frac{1}{L} a^{0c_0}, \quad
f^c[A] = A^{1c} \quad (c \neq c_0),
\label{gaugecond}
\eeq
where $a^0$ denotes the diagonal part of the spatial zero mode of $A^0$.
Note that we cannot require the zero mode of $A^{1c_0}$ to vanish because
of the periodic boundary conditions.  Instead we use the remaining
gauge freedom to eliminate $a^0$.

Now we use (\ref{gaugecond}) to introduce the integration over $a^{0}$
and the missing modes of $A^1$ in (\ref{cont.PI}),
\beq
Z(\beta) \propto \int {\cal D}a^0 \sum_{\omega} \sgn\omega
\int\limits_{\hbox to0pt{\hss$\scriptstyle A^1(\beta)={}^{\omega}\!A^1(0) \atop
\scriptstyle a^1(0) \in {\rm FD}$\hss}}
{\cal D}A^1 \, \delta[f[A]] 
\exp \left[ -\int_0^{\beta} \D \tau \int \D x \left(\frac{1}{2}
(\dot{A}^1)^2 - \frac{V_{\rm eff}}{L} \right)\right],
\label{noncov.Faddeev}
\eeq
where we have suppressed $\det\partial_1$.  $A^1$ contains
$a^1$ as a Cartesian zero mode, and the two integrals
in (\ref{cont.PI}) are contained in the integral over $A^1$.
In the shorthand notation $^{\omega}\!A^1$ the transformation $\omega$
is understood to act only on the diagonal part of the spatial zero mode
of $A^1$.
In the exponent $a^1$ has been replaced with $A^1$ due to the gauge
condition incorporated by the $\delta$-function.

As the next step we generate the remaining modes of the time component of
the gauge field via the identity
\beq
1 = \det {\rm D'_1} \int {\cal D}'A^0 \, \exp \left[ -\int_0^\beta \D \tau
\int \D x
\left( \frac{1}{2} \left( {\rm D_1}A^0 \right)^2 + \dot{A}^1 {\rm D_1}A^0
\right) \right], \label{completing squares}
\eeq
which is valid under the gauge condition. The symbol ${\cal D}'A^0$
means that the zero modes of the covariant derivative D$_1$ are
not included in the integral.

Next we want to show that the Faddeev-Popov determinant is already
contained in the factors appearing in (\ref{noncov.Faddeev}) and
(\ref{completing squares}). It can be calculated by looking at the
change of the the gauge condition under an infinitesimal gauge
transformation,
\beq
\Delta_f [A] = \det \left(\frac{\delta f}{\delta \beta}\right)_{\beta
= 0}.
\eeq
One obtains
\beq
\delta f^{c_0} = \frac{1}{g} \partial_1 \left({\rm D}^1 \delta
\beta \right)^{c_0} + \frac{1}{gL^2}\int\D x \left( {\rm
D}^0\delta\beta \right)^{c_0} , \quad
\delta f^c = \frac{1}{g} \left({\rm D}^1
\delta \beta \right)^c \quad (c \neq c_0),
\eeq
and therefore
\beq
\Delta_f [A] \propto \det {\rm D}^{1\prime},
\eeq
which is, up to a constant factor, what we found above.

The full expression therefore reads
\beq
Z(\beta) \propto \sum_{\omega} \sgn\omega
\int\limits_{\hbox to0pt{\hss$\scriptstyle A^1(\beta)={}^{\omega}\!A^1(0) \atop
\scriptstyle a^1(0) \in {\rm FD}$\hss}}
{\cal D}A^{\mu} \, \delta[f[A]] \, \Delta_f [A]
\exp \left[ -\int_0^{\beta} \D \tau \int \D x \left( \frac{1}{2}
\left(\dot{A}^1 + {\rm D_1}A^0\right)^2 - \frac{V_{\rm eff}}{L} \right)
\right] .
\label{FP-PI}
\eeq
The above expression differs from the traditional QCD path integral by
the sum over residual symmetry transformations and the sign factors.
It should be noted that these features were essential for obtaining
the correct spectrum of pure Yang-Mills theory in section \ref{explicit}.

\section{Discussion}

In this paper we have derived the path integral for 1+1-dimensional QCD
starting from canonically quantized QCD in the gauge fixed formulation
of Lenz et~al.~\cite{LNT}. After eliminating the unphysical degrees of
freedom with the help of Gauss's law one is left with a set of curvilinear
coordinates and the corresponding Jacobian. This yields a non-trivial
form of the kinetic energy which can be simplified by a procedure
analogous to the transition to reduced wave functions in the
canonical formalism. Depending on the domain of definition (compact
vs.\ non-compact) of the remnants of the gauge fields one
can derive different expressions for the transition amplitude.
We compared the two formulations and showed that they are equivalent in
the limit of continuous time. We calculated the
transition amplitude, the partition function and the spectrum in the
pure Yang-Mills theory and found that the latter agrees with the results
of part of the literature \cite{Rajeev,Gupta,Hetrick,Engelhardt,Schreiber}
while it disagrees with some other work \cite{Hetrick,Hosotani}.
The connection to the more standard Faddeev-Popov formulation could 
be established by introducing integrations over the missing gauge
fields. Remainders of the quantum mechanical gauge fixing procedure still
appearing in the Faddeev-Popov expression are a sum over residual
symmetries supplied with corresponding sign factors, and an effective
potential which is, however, constant in this gauge. 

The features which are new to our approach are the direct derivation
of the path integral from the gauge fixed Hamiltonian instead of the
{\em ad hoc\/} introduction of gauge fixing terms; the notorious
``Faddeev-Popov infinities'' are avoided.  In our construction
residual symmetries and boundary conditions on the wave functions are
explicitly taken care of.

The next steps in this direction would be to work out a corresponding
path integral expression for 3+1-dimensional QCD\@. There are problems
arising from the fact that the path-ordered exponential depends on
two space coordinates.

We hope that the connection between the canonical and path integral
approaches provided by our calculation can contribute to a better
understanding of the renormalization of the canonical formulation of QCD\@.
Furthermore, our work might help to clarify the relevance of the
Jacobian. Because we are using a manifestly gauge invariant approach 
our formulation of QCD is well-suited for all kinds of approximations
without the risk of violating gauge invariance. One might hope to
find traces of non-perturbative physics relevant for confinement in
3+1 dimensions in this way.

\section{Acknowledgements}

We are indebted to Prof.~F.~Lenz for help and advice during the course of
this work. We further gratefully acknowledge discussions with
Prof.~L.~O'Raifeartaigh, Prof.~H.~Reinhardt, Prof.~M.~Thies and
Dr.~H.~W.~Grie{\ss}hammer.

\appendix

\section{Inclusion of Fermions}

In this appendix we derive the transition amplitude for the case of
fundamental quarks coupling to the gauge field.

The physical Hamiltonian in normal-ordered form (i.e.\ all
$\psi^\dagger$ to the left of all $\psi$) reads, after the unitary
transformation in \cite{LNT},
\beq
H_{\rm ph} = H_{\rm F} + H_{\rm G} + H_{\rm C},
\eeq
where
\beq
H_{\rm F} = \int \D x \, \psi^\dagger \left(\alpha \left(
\frac{\partial}{\I \partial x} - g a \right) + \beta m \right) \psi, 
\eeq
and
\beqa
\lefteqn{
H_{\rm C} = \frac{g^2}{2} \int \D x \int \D y \, \psi^\dagger_i (x)
\frac{\lambda_{ij}^a}{2} \psi^\dagger_k (y) \left[ \frac{1}{- {\rm
d}'^2} \right]^{ab} \!\!\!\! (x-y) \, \psi_j (x)
\frac{\lambda_{kl}^b}{2} \psi_l (y)
+{} } \nonumber \\ & & \qquad\qquad
{}+ \frac{g^2L}{8\Nc} \left[ \frac{\Nc^2-1}{6\Nc} 
+ \smash{\sum_{k>l}} \sin^{-2} \left(\frac{gL(a_l-a_k)}{2}\right)
\right] Q_{\rm m} 
\eeqa
is the non-Abelian Coulomb interaction.  Here, the $\psi$ are
gauge invariant fermion operators, and the form
of $H_{\rm G}$ is that appearing in (\ref{yangmillsphysical}).  d$'$
is the covariant derivative in the adjoint representation
corresponding to $a$, i.e.\ $\partial/\partial x-\I g[a,.\,]$, on the space
of its non-zero modes.  The contraction arising in the Coulomb term
gives, apart from the term proportional to the electromagnetic
fermionic charge $Q_{\rm m}$, a term which is proportional to the
diagonal part of the chromo-electric fermionic charge and vanishes on
physical states because of the covariant zero mode of Gauss's law:
\beq
Q_{\rm m}^{c_0} | \tilde\Phi\rangle = 0,
\eeq
where $| \tilde\Phi\rangle$ denotes a physical state after the unitary
transformation in \cite{LNT}.

We now derive a path integral expression for the amplitude $\langle
a_{\rm f}, \psi_{\rm f}| \,\E^{-\I H_{\rm ph} t} | \tilde\Phi \rangle$,
where $|\psi\rangle$ denotes a coherent state for a Dirac spinor.

The time-slicing procedure is made possible by the identity \cite{Negele}
\beq
\E^{-\I H t} = \lim_{N \rightarrow \infty} \left[\mathopen: \E^{-\I H t/N}
\mathclose:\right]^N,
\eeq
valid for normal-ordered Hamiltonians $H$.

In the ``time-sliced'' transition function we insert at each step the
identity in terms of coherent states of Dirac spinors \cite{Negele},
\beq
1 = \int \D \psi^\dagger \D \psi \, \exp \left[ - \int \D x \,
\psi^\dagger \psi \right] | \psi \rangle \langle \psi |,
\eeq
as well as the identity expressed by eigenstates of gauge-field
operators (see (\ref{amplitude})). We then evaluate all the operators in
a procedure similar to the above. The final result is
\beqa
\lefteqn{\langle a_N, \psi_N| \,\E^{-\I H_{\rm ph} t} | \tilde\Phi \rangle 
= \left( \frac{L}{2\pi \I \varepsilon} \right)^{(N_{\rm c}-1)N/2} 
\int \prod_{i=0}^{N-1} \left[ \D a_i \D \psi^\dagger_i \D \psi_i \right]  
\sqrt{\frac{J(a_0)}{J(a_N)}} \,
\frac{\sgn a_0}{\sgn a_N}\times{}}
\nonumber\\
&& {}\times\exp \left[ \I \varepsilon \sum_{i=1}^N \left(
\frac{L}{2}\frac{(a_{i}-a_{i-1})^2}{\varepsilon^2} - V_{\rm eff} 
- \int \D x \, \I \frac{\psi^\dagger_i - \psi^\dagger_{i-1}}{\varepsilon}
\psi_{i-1} - H_{\rm F} - H_{\rm C}
\right)\right]\times{}
\nonumber\\
&& {}\times\langle a_0, \psi_0| \tilde\Phi \rangle. \qquad
\eeqa
In this formula $H_{\rm F}$ and $H_{\rm C}$ depend on the fields
$\psi^\dagger_i$, $\psi_{i-1}$ and $a_{i-1}$.

\end{document}